\RequirePackage{heppennames2}
\RequirePackage{lhchiggs}
\documentclass[12pt,a4paper]{article}
\usepackage{graphicx} 
\usepackage{graphicx}
\usepackage{tabularx}
\usepackage{subcaption}
\usepackage{placeins}
\usepackage{multirow}
\usepackage{graphbox}
\usepackage{hyperref}
\usepackage{color}
\usepackage{cite}
\usepackage[normalem]{ulem}
\begin{document}

\title{
\vspace*{-1cm}
\begin{flushright}
{\small LHCHWG-2025-008} \\[0.5cm]
\end{flushright}
Higgs-Boson Decays: Updates}
\author{Emanuele Bagnaschi$^1$, Lisa Biermann$^2$ and Michael Spira$^2$}
\date{}

\maketitle
\noindent
{\it $^1$ INFN, Laboratori Nazionali di Frascati, Via E. Fermi 40, 00044 Frascati (RM), Italy \\
$^2$ PSI Center for Neutron and Muon Sciences, 5232 Villigen PSI, Switzerland}

\abstract{\noindent In this contribution, new developments for the Standard Model Higgs-boson decays will be summarized. This addresses extensions of the grids used by {\tt Hdecay} to implement finite NLO mass effects for $H\to gg$ as well as explicit numbers for the branching fraction for the strange-Yukawa induced part of $H\to s\bar s$ together with the related uncertainties. In addition, first results of the strong and weak Dalitz decays of the Higgs-boson decays $H\to s\bar s + g/\gamma$ are presented. The Dalitz decays define the separation between the Yukawa-induced part and the continuum, that is not induced by the strange-Yukawa coupling, and thus pave the way towards a solid determination of the strange-Yukawa coupling at future $e^+e^-$ colliders, but will also be relevant for solid bounds on the strange-Yukawa coupling at the LHC.}

\section{Introduction}
The discovery of the Higgs boson \cite{Higgs:1964ia, Higgs:1964pj,
Englert:1964et, Guralnik:1964eu, Higgs:1966ev, Kibble:1967sv} with a mass of 125 GeV at the LHC
\cite{Aad:2012tfa,Chatrchyan:2012xdj} completed
the SM of strong and electroweak interactions provided it is compatible with the Standard Model (SM) Higgs boson. However, the consistency
with the SM Higgs boson needs to be probed in more
detail by measuring the couplings to other SM particles. This is
pursued by deriving the related couplings from Higgs-boson
production and decays at the LHC \cite{Khachatryan:2016vau,ATLAS:2019nkf,CMS:2020xwi} and
will be continued in the future runs. The extraction of the SM
parameters from the observables is plagued by experimental and
theoretical uncertainties that have to be determined consistently.
The determination of the branching ratios of Higgs boson decays thus
necessitates the inclusion of the available higher-order corrections
(for an overview see e.g.~Ref. \cite{Spira:2016ztx}) and a
sophisticated estimate of the theoretical and parametric uncertainties.

In this contribution to the LHC Higgs Working Group Report 5, this work will address the improvements of $H\to gg$ related to the implementation of finite quark-mass effects at NLO QCD beyond Higgs masses of 1 TeV in Section \ref{sc:h2gg}, provide predictions of the Yukawa-induced contribution to $H\to s\bar s$ in Section \ref{sc:h2ss0} and finally discuss strong and weak Dalitz decays in Section \ref{sc:dalitz} with particular emphasis on the Higgs decays into strange quarks, while ending up with short conclusions in Section \ref{sc:conclusions}.

\section{$H\to gg$} \label{sc:h2gg}
\begin{figure}[hbt]
\vspace*{-4.5cm}

    \centering
    \includegraphics[width=0.7\textwidth]{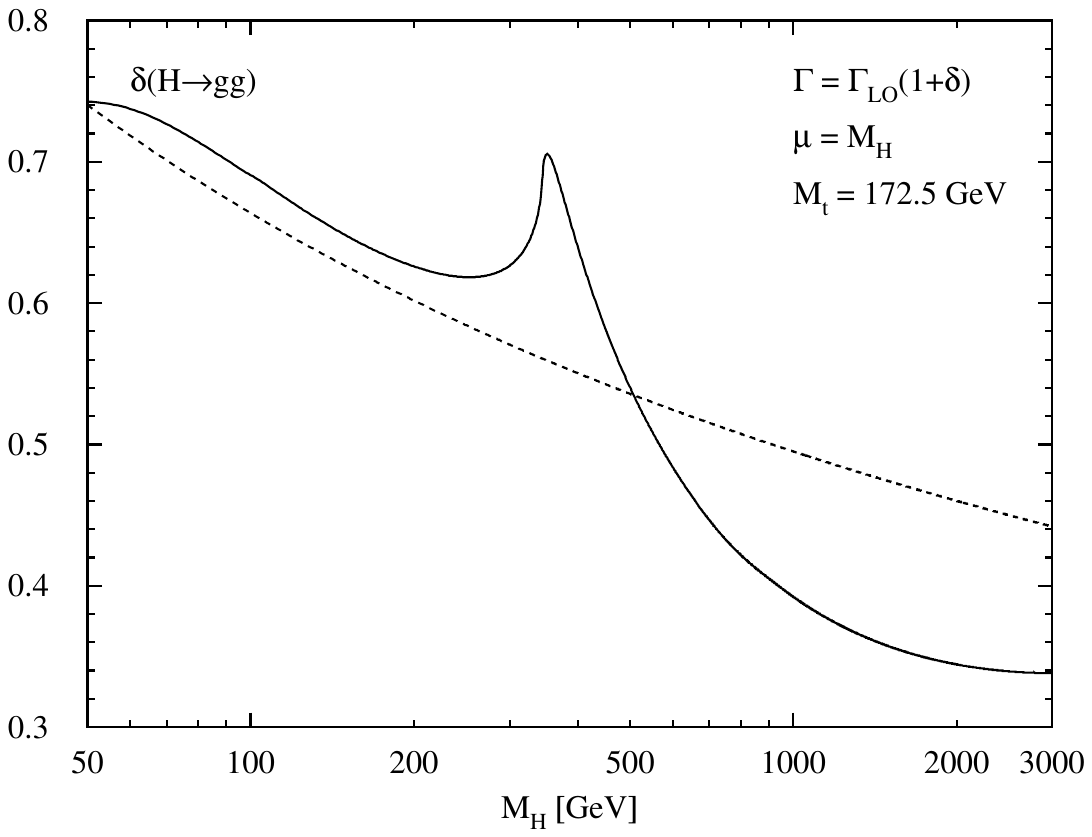}
    \vspace*{-4.5cm}
    
    \caption{\it Relative NLO QCD corrections with and without finite quark mass effects. The dashed line shows the NLO QCD corrections in the heavy-quark limit (HQL) for all quarks, while the full line includes the full quark-mass effects at NLO QCD. The corrections are related to the notation of Eq.~(\ref{eq:h2ggqcd}) by $\delta = E~\alpha_s/\pi$, where $\Delta E$ is put to zero in the HQL.}
    \label{fg:h2gg}
\end{figure}

The loop-induced Higgs decay into gluons reaches a branching ratio of
about 8\%. The decay is dominantly mediated by top and bottom quark
loops, with the latter providing a 10\% contribution to this partial decay width, while the charm quark contributes at the level of about 2\%. The two-loop QCD corrections are
known, including the exact quark mass dependencies \cite{Inami:1982xt,
Djouadi:1991tka, Spira:1995rr},
\begin{equation}
\Gamma(H\rightarrow gg(g),~gq\bar q) = \Gamma_{LO}(H\rightarrow gg)
\left[ 1 + E \frac{\alpha_s}{\pi} \right]
\end{equation}
with ($N_F=5$)
\begin{equation}
E = \frac{95}{4} - \frac{7}{6} N_F
+ \frac{33-2N_F}{6}\ \log \frac{\mu^2}{M_H^2} + \Delta E
\label{eq:h2ggqcd}
\end{equation}
where $\Gamma_{LO}(H\rightarrow gg)$ denotes the gluonic decay width at LO, $\alpha_s$ the strong coupling, $\mu$ the renormalization scale, and $\Delta E$ denotes the finite quark-mass effects beyond the heavy quark limit (HQL) at NLO [normalized to the fully mass-dependent LO decay width $\Gamma_{LO}(H\rightarrow gg)$]. The NLO QCD corrections enhance the partial decay width by
about 70\%. The NNLO, N$^3$LO, and the N$^4$LO QCD corrections
have been obtained for the top loops in the limit of heavy top quarks,
i.e.~the leading term of a large top mass expansion
\cite{Chetyrkin:1997iv, Baikov:2006ch, Herzog:2017dtz}. The QCD
corrections beyond NLO amount to less than 20\% of the NLO QCD-corrected
partial decay width, thus underlining perturbative convergence in spite of
the large NLO corrections. The residual theoretical uncertainties have
been estimated at about 3\% from the scale dependence of
the QCD-corrected partial decay width. The NLO electroweak (elw.) corrections have
been obtained for the top-loop contributions first in the limit of
heavy top quarks \cite{Djouadi:1994ge, Chetyrkin:1996wr,
Chetyrkin:1996ke}, then the electroweak corrections involving light
fermion loops exactly \cite{Aglietti:2004nj, Aglietti:2006yd,
Degrassi:2004mx}, and finally the full electroweak corrections involving
$W, Z$, and top-loop contributions, including the full virtual mass
dependencies, by means of a numerical integration \cite{Actis:2008ug,
Actis:2008ts}. They amount to about 5\% for the SM Higgs mass value.
The public tool {\tt Hdecay} \cite{Djouadi:1997yw, Djouadi:2018xqq}
includes the NLO QCD results with the full quark mass dependencies, the
NNLO and N$^3$LO QCD corrections in the heavy top limit (HTL), and the full NLO
electroweak corrections in terms of a grid in the Higgs and top masses used for
an interpolation. The grids for the top and bottom quark-mass effects $\Delta E$ of Eq.~(\ref{eq:h2ggqcd}) in the NLO QCD corrections have now been extended to a Higgs mass of 3 TeV for BSM studies in the code {\tt Hdecay} \cite{Djouadi:1997yw, Djouadi:2018xqq}, see Fig.~\ref{fg:h2gg}. Since the charm contribution is small overall, its NLO mass effects beyond the HQL are negligible.  This extension is a new development after Yellow Report 4 (YR4) \cite{deFlorian:2016spz} and available in {\tt Hdecay}.

\section{$H\to f\bar f$} \label{sc:h2ss0}
The Higgs decay $H \to b\bar b$ is the dominant Higgs boson decay with a
branching ratio of about 58\%. The subleading fermionic decays $H\to
\tau^+\tau^-$ and $\mathrm{H}\to \mathrm{c\bar c}$ reach branching ratios of about 6\% and
3\%, respectively. The rare decay $\mathrm{H} \to \mu^+\mu^-$ will become visible
at the HL--LHC and happens with about 0.02\% probability
\cite{deFlorian:2016spz}. The present status of the partial decay
widths can be summarized in terms of the (factorized) expression
\begin{equation}
\Gamma (H\to f{\overline{f}}) = \frac{N_c G_F M_H } {4\sqrt{2}\pi}
\ m_f^2\ (1 + \delta_{QCD}+\delta_t + \delta_{mixed})
\left(1+\delta_{elw} \right) \, ,
\end{equation}
where $N_c=3(1)$ for quarks (leptons), $G_F$ denotes the Fermi constant,
$M_H$ denotes the Higgs mass, and $m_f$ denotes the fermion mass. The QCD
corrections $\delta_{QCD}$ to the Higgs boson decays into quarks are
known up to NLO including the full quark mass dependence
\cite{Braaten:1980yq, Sakai:1980fa, Inami:1980qp, Drees:1990dq,
Drees:1989du} and up to N$^4$LO for the leading corrections with the
leading mass effects \cite{Gorishnii:1983cu, Gorishnii:1990zu,
Gorishnii:1991zr, Kataev:1993be, Surguladze:1994gc, Chetyrkin:1996sr,
Melnikov:1995yp}. The dominant part of the QCD corrections can be
absorbed in the running quark mass evaluated at the scale of the Higgs
mass. The top-induced QCD corrections $\delta_t$, which are related to interference
effects between $H\to gg$ and $H\to q\bar q$, are known at NNLO in the
limit of heavy top quarks and light bottom quarks
\cite{Chetyrkin:1995pd, Larin:1995sq, Primo:2018zby} and more recently including the full mass dependence \cite{Bernreuther:2018ynm,Behring:2019oci,Somogyi:2020mmk,Wang:2023xud} with first results even at N$^3$LO \cite{Wang:2024ilc}. In the case of
leptons, there are no QCD corrections ($\delta_{QCD}=\delta_t =
\delta_{mixed}=0$). The electroweak corrections $\delta_{elw}$ are known
at NLO exactly \cite{Fleischer:1980ub, Bardin:1990zj, Dabelstein:1991ky,
Kniehl:1991ze}. The mixed QCD-electroweak~corrections $\delta_{mixed}$ rank at the
per-mille level if the factorized expression with respect to QCD and
electroweak corrections is used \cite{Kataev:1997cq, Kniehl:1994ju,
Kwiatkowski:1994cu, Chetyrkin:1996wr, Mihaila:2015lwa, Chaubey:2019lum}.
The public tool {\tt Hdecay} \cite{Djouadi:1997yw, Djouadi:2018xqq}
neglects the mixed QCD-electroweak corrections but includes all other
corrections.

The parametric errors are dominated by the uncertainties in the top,
bottom, and charm quark masses, as well as the strong coupling $\alpha_\mathrm{s}$.
We have used the $\overline{\rm MS}$ masses for the bottom and charm
quarks, $\overline{m}_b (\overline{m}_b) = (4.18 \pm 0.03)~\mbox{GeV}$ and
$\overline{m}_c (3~\mbox{GeV}) = (0.986 \pm 0.026)~\mbox{GeV}$, and the
top quark pole mass $m_t = (172.5 \pm 1)~\mbox{GeV}$, according to the
conventions of the LHC Higgs Working Group (LHCHWG)
\cite{deFlorian:2016spz}. The $\overline{\rm MS}$ bottom and charm
masses are evolved from the input scale to the scale of the decay
process with three-loop accuracy in QCD. The strong coupling $\alpha_\mathrm{s}$ is
fixed by the input value at the Z boson mass scale, $\alpha_\mathrm{s}(M_\mathrm{Z}) =
0.118 \pm 0.0015$.  The total parametric uncertainty for each branching
ratio has been derived from a quadratic sum of the individual impacts of
the input parameters on the decay modes along the lines of the original
analyses in Refs.~\cite{Djouadi:1995gt, Gross:1994fv} and the later analysis
in Ref.~\cite{Denner:2011mq}. The present recommended numbers of the LHCHWG for the partial decay widths and branching ratios have been obtained using {\tt Prophecy4f} \cite{Bredenstein:2006rh,
Bredenstein:2006ha} for the decays $H\to WW,ZZ$ and {\tt Hdecay}
\cite{Djouadi:1997yw, Djouadi:2018xqq} for the other decay modes.

As a new ingredient after YR4 \cite{deFlorian:2016spz}, we are providing theoretical predictions and uncertainties of the Yukawa-induced part of the Higgs-boson decay into strange quarks.
Adopting the value $\overline{m}_s (2~\mbox{GeV}) = (93.5 \pm 3.2)~\mbox{MeV}$ for the strange $\overline{MS}$ mass \cite{ParticleDataGroup:2024cfk}\footnote{We have enhanced the error by a factor of four to be conservative.}, we have obtained the Yukawa-induced branching ratios of $H\to s\bar s$ with the corresponding theoretical and parametric uncertainties. For a selection of Higgs masses they can be found in Tab.~\ref{tb:h2ss}. These values serve as reference points for the extraction of the strange-Yukawa coupling (see Section \ref{sc:h2ss}) and for studies beyond the SM.
\begin{table}[hbt]
\centering
\renewcommand{\arraystretch}{1.5}
\begin{tabular}{|l|llll|} \hline
$M_H$ [GeV] & $BR(H\to s\bar s)$ & THU [\%] & PU($m_q$) [\%] & PU($\alpha_s$) [\%] \\ \hline
120    & $2.380 \cdot 10^{-4}$ & +0.73 --0.73 & +7.03 --6.80 & +1.98 --2.02 \\
121    & $2.332 \cdot 10^{-4}$ & +0.73 --0.73 & +7.03 --6.80 & +1.98 --2.06 \\
122    & $2.282 \cdot 10^{-4}$ & +0.73 --0.73 & +7.03 --6.79 & +2.03 --2.07 \\
123    & $2.230 \cdot 10^{-4}$ & +0.73 --0.73 & +7.03 --6.79 & +2.05 --2.07 \\
124    & $2.175 \cdot 10^{-4}$ & +0.73 --0.73 & +7.02 --6.79 & +2.04 --2.13 \\
125    & $2.119 \cdot 10^{-4}$ & +0.73 --0.73 & +7.02 --6.78 & +2.10 --2.10 \\
125.09 & $2.114 \cdot 10^{-4}$ & +0.73 --0.73 & +7.02 --6.78 & +2.11 --2.10 \\
126    & $2.061 \cdot 10^{-4}$ & +0.73 --0.73 & +7.01 --6.78 & +2.09 --2.16 \\
127    & $2.001 \cdot 10^{-4}$ & +0.73 --0.73 & +7.01 --6.78 & +2.09 --2.18 \\
128    & $1.940 \cdot 10^{-4}$ & +0.73 --0.73 & +7.00 --6.77 & +2.17 --2.16 \\
129    & $1.877 \cdot 10^{-4}$ & +0.73 --0.73 & +7.00 --6.77 & +2.19 --2.20 \\
130    & $1.813 \cdot 10^{-4}$ & +0.73 --0.73 & +6.99 --6.77 & +2.18 --2.27 \\ \hline
\end{tabular}
\caption{\it Branching ratios for the strange-Yukawa induced contribution to $H\to s\bar s$ in the Higgs-mass range between 120 and 130 GeV and the individual uncertainties, i.e.~theoretical (THU) and parametric due to the quark masses [PU($m_q$)] and $\alpha_s$ [PU($\alpha_s$)].}
\label{tb:h2ss}
\end{table}

\section{$H\to Z\gamma$ and Dalitz decays} \label{sc:dalitz}
The rare loop-induced Higgs decay into a Z boson and a photon reaches a
branching ratio of less than 0.2\%. The decay is mediated by W
and top quark loops dominantly, with the W loops being leading. The two-loop QCD
corrections are known, including the exact top mass dependencies
\cite{Spira:1991tj, Bonciani:2015eua, Gehrmann:2015dua}. They correct
the partial decay width at the per-mille level and
thus can safely be neglected. The electroweak corrections to this decay
mode have recently been calculated and shown to be small (i.e.~below the 1\%-level) once the partial decay width is defined in the $G_F$ scheme for the couplings to the Higgs and $Z$ boson and using $\alpha(0)$ in the Thomson limit for the photonic coupling \cite{Chen:2024vyn,Sang:2024vqk}. However, the decay mode $H\to Z\gamma\to f\bar
f\gamma$ is a subset of the more general Dalitz decays $H\to f\bar f\gamma$
\cite{Abbasabadi:1996ze, Abbasabadi:2006dd, Abbasabadi:2004wq,
Dicus:2013ycd, Chen:2012ju, Passarino:2013nka, Sun:2013rqa, Kachanovich:2020xyg, Kachanovich:2021pvx}. The latter are described by the diagrams in Fig.~\ref{fg:h2ss_dia}, where the
$Z$ boson exchange appears in a subset of the triangle diagrams. The resonant $Z$ boson exchange corresponds to the $H\to Z\gamma$ decay
mode. The separation of this part, however, depends on the experimental
strategy to reconstruct the $Z$ boson in the final state. The issue of Higgs Dalitz decays is entirely an ingredient beyond the studies of YR4 \cite{deFlorian:2016spz}.

\subsection{Dalitz decays into strange quarks} \label{sc:h2ss}
The measurement of the Yukawa coupling to strange quarks is one of the physics goals
of the $e^+e^-$ Higgs factories that will possibly be operational in the second part of the century.
Recent studies performed by the ECFA show quite promising possibilities\cite{Altmann:2025feg}.
However, the proper definition and simulation of the signal and background processes is highly relevant.

\begin{figure}[hbt]
\vspace*{-2cm}

    \centering
    \includegraphics[width=1.2\textwidth]{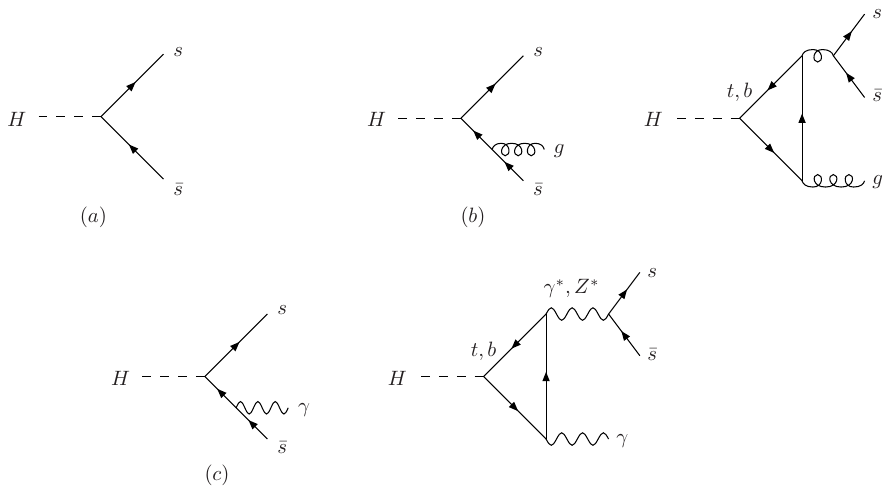}
    \vspace*{-19cm}
    
    \caption{\it Typical diagrams contributing to (a) $H\to s\bar s$, (b) $H\to s\bar s g$ and (c) $H\to s\bar s\gamma$.}
    \label{fg:h2ss_dia}
\end{figure}
The strange-Yukawa coupling induces Higgs decays into strange quarks as shown in Fig.~\ref{fg:h2ss_dia}a. The LO process is only moderately modified by QCD and electroweak corrections, provided the strange-Yukawa coupling is defined in terms of the $\overline{MS}$ strange mass $\overline{m}_s(M_{H})$ at the scale of the Higgs mass $M_{H}$\footnote{The running strange mass amounts to $\overline{m}_s (M_H) = (52.9 \pm 1.8)$ MeV with 3-loop running for the Higgs mass of $M_H = 125$ GeV.}. The inclusive branching ratio of the strange-Yukawa induced part, $H\to s\bar s$, amounts to $0.021\%$ (see Table \ref{tb:h2ss}). Thus, it is very small, i.e.~below the per-mille level. Therefore, competing Higgs decays with final-state strange quarks not induced by the strange-Yukawa coupling are relevant. The dominant decays of this type are the strong and weak Higgs Dalitz decays, $H\to s\bar s + g/\gamma$, see Fig.~\ref{fg:h2ss_dia}b,c, that are loop-induced. Their inclusive contribution to the total branching ratios ranks at the per-cent level, so that they are more than an order of magnitude larger than the branching ratio of $H\to s\bar s$ induced by the strange-Yukawa coupling. In order to disentangle the different contributions to facilitate a measurement of the strange-Yukawa coupling, however, the exclusive distributions have to be investigated in detail, so that suitable cuts on e.g.~the final-state $s\bar s$-invariant mass can be found that will have a different impact on the loop-induced and Yukawa-induced parts.

\begin{figure}[hbt]
    \vspace*{-3.7cm}
    
        \centering
        \hspace*{-0.5cm}
        \includegraphics[width=0.5\textwidth]{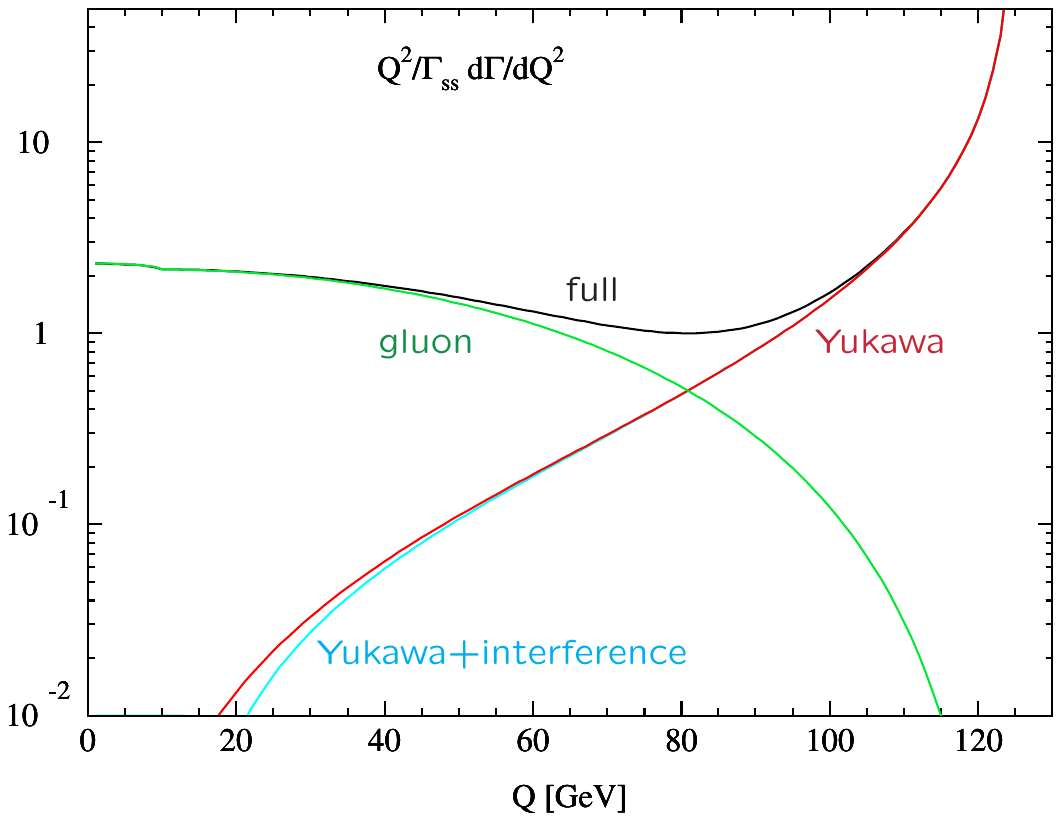} \hspace*{0.0cm}
        \includegraphics[clip=true, width=0.495\textwidth, trim={1.5cm 1.0cm -1.4cm 0.0cm}]{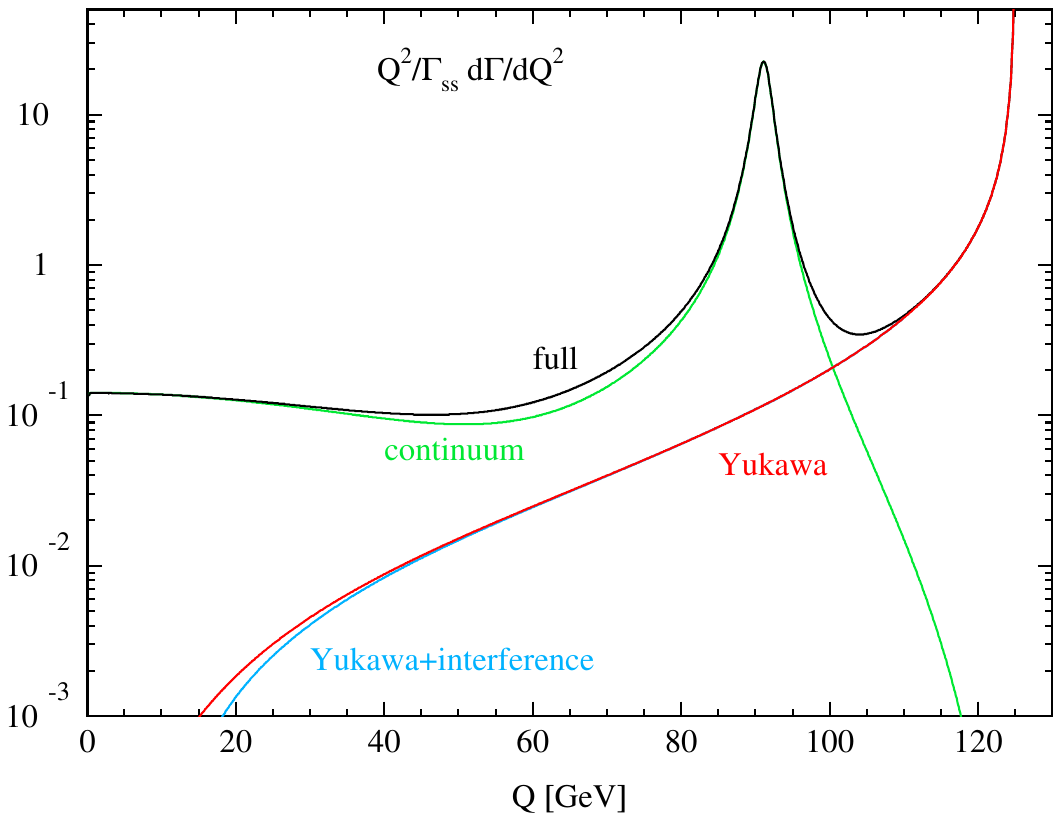} \hspace*{-1.0cm}
        \vspace*{-2.7cm}

        \hspace*{-0.6cm} (a) \hspace*{7.2cm} (b)
    \caption{\it The differential partial width of (a) the strong Dalitz decay $H\to s\bar s g$ and (b) the weak Dalitz decay $H\to s\bar s\gamma$ as a function of the invariant $s\bar s$ mass $Q$ normalized to the inclusive partial decay width $\Gamma_{ss} = \Gamma(H\to s\bar s)$. The kinematical strange-mass has been set to zero. No resummation of soft gluon and photon effects has been performed at the upper end of the spectra, while finite strange-mass effects will be relevant only in the low-$Q$ region.}
    \label{fg:dalitz}
\end{figure}

In our analysis, we have studied the strong and weak Dalitz decays $H\to s\bar s + g/\gamma$ including the interference with the Yukawa-induced part. The differential strong Dalitz decay width for $H\to s\bar s g$ is depicted in Fig.~\ref{fg:dalitz}a as a function of the invariant $s\bar s$ mass $Q=M_{s\bar s}$. The separate parts of the loop-induced (gluon) contribution, the Yukawa-induced one and the interference are presented as well as the total sum. If the mass resolution of the final-state strange jet pair is sufficient, the range of large $Q$ values could be extracted so that the strange-Yukawa coupling could be constrained or measured. The differential decay width of Fig.~\ref{fg:dalitz}a is multiplied by $Q^2$ and normalized to the total partial width $\Gamma(H\to s\bar s)$\footnote{The finite limit of the loop-induced part at low values of $Q$ underlines the proper collinear behaviour of this $Q^2$-multiplied distribution related to the final-state $s\bar s$ pair.}.

The weak Dalitz decay $H\to s\bar s + \gamma$ has been calculated analogously. The resulting partial decay width involving all loop-induced and strange-Yukawa induced contributions is presented in Fig.~\ref{fg:dalitz}b. Due to the contributing $Z$-boson resonance, the differential weak Dalitz decay width is of similar size as the strong Dalitz decay width. As in the case of the strong Dalitz decay, the large-$Q$ range of the distribution is dominated by the contribution induced by the Yukawa coupling. Thus, the invariant strange-jet pair mass resolution plays the crucial role for the sensitivity to the strange-Yukawa coupling.

There are open questions related to this sensitivity, i.e.~the detailed definition of the strange mass at the jet level related to the definition of the strange mass contained in the Yukawa coupling. Moreover, fragmentation effects of the more dominant $b\bar b, c\bar c$ Higgs-boson decays into strange jets matter in this context.

\section{Conclusions} \label{sc:conclusions}
In this work new developments of the SM Higgs branching ratios beyond YR4 have been summarized. Particular emphasis has been set on Higgs Dalitz decays and in particular the strange-Yukawa coupling in these processes with strange quarks in the final state. Moreover, an extension of the grids for NLO mass effects on $H\to gg$ has been provided that allows for more extensive studies in BSM frameworks.

~\\
\noindent
{\bf Acknowledgements} \\
We are grateful to S.~Heinemeyer, A.~M\"uck, I.~Pulak and D.~Rebuzzi for useful comments on this work. The work of L.B. is supported by the Swiss National Science Foundation (SNSF).

\bibliography{br}

\end{document}